\documentclass[letterpaper,prc,twocolumn,showpacs,floatfix,nofootinbib,preprintnumbers,superscriptaddress,amsmath,amssymblinenumbers]{revtex4-2} %

\usepackage{hyperref}
\usepackage{amsmath}
\usepackage{amsfonts}
\usepackage{graphicx}
\usepackage{hyperref}
\hypersetup{
    colorlinks=true,  %
    linkcolor=blue,
    urlcolor=blue,
    citecolor=blue
}

\newcommand{\diminp}{n_{d}}
\newcommand{\dimout}{p}
\newcommand{\dimsamp}{n}
\newcommand{\ntotal}{n}
\newcommand{\ntrain}{n_t}
\newcommand{\nvalid}{n_v}

\newcommand{\mlinp}{\boldsymbol{X}}
\newcommand{\mlout}{\boldsymbol{Y}}
\newcommand{\anchors}{\boldsymbol{C}}
\newcommand{\loss}{\mathcal{L}}
\newcommand{\duq}{$\Delta$-UQ}
\usepackage{stix}

\begin{document}

\preprint{LLNL-JRNL-2004673}

  \title{Quantifying uncertainty in machine learning on nuclear binding energy}
  \author{Mengyao Huang}
  \email[Contact author: ]{huang59@llnl.gov}
  \affiliation{Lawrence Livermore National Laboratory, P.O.\ Box 808, L-414, Livermore, California 94551, USA}
  \author{Kyle A.\ Wendt}
  \affiliation{Lawrence Livermore National Laboratory, P.O.\ Box 808, L-414, Livermore, California 94551, USA}
  \author{Nicolas F.\ Schunck}
  \affiliation{Lawrence Livermore National Laboratory, P.O.\ Box 808, L-414, Livermore, California 94551, USA}
  \author{Erika M.\ Holmbeck}
  \affiliation{Lawrence Livermore National Laboratory, P.O.\ Box 808, L-414, Livermore, California 94551, USA}

  \begin{abstract}
Techniques from artificial intelligence and machine learning are increasingly employed in nuclear theory, however, the uncertainties that arise from the complex parameter manifold encoded by the neural networks are often overlooked. Epistemic uncertainties arising from training the same network multiple times for an ensemble of initial weight sets offer a first insight into the confidence of machine learning predictions, but they often come with a high computational cost.
Instead, we apply a single-model uncertainty quantification method called {\duq} that gives epistemic uncertainties with one-time training. We demonstrate our approach on a 2-feature model of nuclear binding energies per nucleon with proton and neutron number pairs as inputs.
We show that {\duq} can produce reliable and self-consistent epistemic uncertainty estimates and can be used to assess the degree of confidence in predictions made with deep neural networks.
\protect\\\\{DOI: \hyperlink{https://doi.org/10.1103/tdk4-c4tp}{10.1103/tdk4-c4tp}}
  \end{abstract}

  \maketitle

\label{sec:intro}

\section{Introduction}

The nuclear binding energy is a fundamental property of atomic nuclei 
and is the key driver of the energy released in nuclear reactions. An accurate and precise description of nuclear binding energy across the entire chart of nuclides is thus an important ingredient in modeling nuclear reactions relevant for medical applications \cite{chandra2017nuclear}, nuclear energy \cite{devanathan2010modeling} and astrophysical studies \cite{wiescher2012critical}. For instance, complex networks of nuclear reactions are involved in neutron star mergers \cite{RN1,RN2} and core-collapsed supernovae \cite{RN3}. However, current nuclear mass tables \cite{Huang_2021,wang2021ame} are neither complete nor sufficiently accurate for these astrophysical applications \cite{neufcourt2020quantified}. High-precision experimental measurement data are available near the valley of nuclear stability; while efforts are being made to continue expanding the measured region towards the nuclear drip lines \cite{Gaulard_MISTRAL02,doi:10.1080/10619127.2017.1317176,refId0}, the nuclear masses of only about 40\% of all nuclei predicted to exist have been measured. Theoretical mass models are either based on semi-empirical macroscopic-microscopic approaches \cite{MOLLER20161,MYERS1996141,royer2010macro}, phenomenological microscopic models \cite{DUFLO199429,ZUKER199465}, or non-relativistic energy density functional theory \cite{ryssens2022skyrme,goriely2013hartreefockbogoliubov,goriely2009first}. Masses predicted by these approaches are consistent within the experimentally measured region, where they are calibrated, but predictions vary significantly in neutron-rich or superheavy nuclei.

Recently, machine learning has become a powerful tool to reproduce several nuclear properties, including nuclear masses \cite{PhysRevC.109.064322,PhysRevC.106.014305,RN5,LI2024138385,PhysRevC.106.L021301}, charge radii \cite{bayram2023applications,DONG2023137726}, and nuclear reaction cross sections \cite{JIN2024111545,RN6}. Unlike traditional methods, which rely on physical models, machine learning methods use neural networks that employ linear and nonlinear layers to directly fit the available data. The existence of apparent patterns and strong trends in nuclear binding energies across the chart of isotopes provides a strong incentive to employ machine learning to learn the correlations responsible for these patterns.
However, extreme caution must be taken when extending predictions far away from the training data points on the nuclear chart. Barring the guidance of a physical model, a reliable uncertainty quantification method is critical in deciding when and to what extent one can trust the machine learning results.

The two main sources of uncertainties are {\em epistemic} uncertainties due to limited data and a lack of knowledge of the best model in the hypothesis space, and {\em aleatoric} uncertainties that are irreducible by increasing data and knowledge. Distinguishing between epistemic and aleatoric uncertainties is hard to do in methods such as Gaussian processes \cite{PhysRevC.109.064322}, Bayesian neural networks \cite{neufcourt2018bayesian,neufcourt2019neutron}, and probabilistic networks \cite{PhysRevC.106.014305} that generate uncertainties from posterior distributions. For example, Ref.~\cite{PhysRevC.106.014305} runs the probabilistic network 50 times for the resulting distribution as an indication of reproducibility, but it is not clear how much of this total uncertainty is caused by epistemic uncertainties alone. By contrast, deterministic networks lend themselves more easily to separating and evaluating epistemic uncertainties, although this has rarely been done in nuclear physics. In Ref.~\cite{RN5}, the authors run the network 500 times with different random split of training and test data to obtain an estimate of them.

Given a neural network with a fixed number of layers and nodes, training the network for a set of random initial conditions allows for exploring the hypothesis space. Since training involves minimizing a function---the loss function---of a potentially large number of variables---the weights of the network---small differences in the initial numerical values of the weights can lead to substantially different solutions. Epistemic uncertainties can thus be estimated by independently training multiple copies of the same neural network with different initial weights and analyzing the spread of the results. This is an example of the ensemble methods \cite{NIPS2017_9ef2ed4b,10.5555/3454287.3455541,rahaman2021uncertainty,valdenegro2023sub} which offer a straightforward approach to estimate epistemic uncertainties. However, the multiple independent runs needed induce a high computational cost.

Single-model methods \cite{RN8,RN9} have been emerging as a more efficient way to estimate epistemic uncertainty. They only require running the network once with a unique set of initializations, and are able to generate both the predictions and epistemic uncertainties through that unique run. {\duq} \cite{thiagarajan2022single} is one of these methods. By combining inputs with different constant biases, {\duq} generates uncertainties resembling those from ensemble methods but reduces total computational resource consumption by a factor of $\sim X$, where $X$ is the size of the ensemble in the ensemble method. {\duq} also provides a convenient way to assess risk levels \cite{pmlr-v235-j-thiagarajan24a} to extend the extrapolation capabilities of usual single-model methods \cite{RN8}.

The goal of this work is to test the effectiveness of the {\duq} uncertainty quantification method in the case of deep neural network models of nuclear binding energy per nucleon ($E/A$). In particular, we show that {\duq} provides a quantitative indicator to assess the reliability of deep neural network predictions in extrapolations far from the training region.

The paper is organized as follows: in Sec.~\ref{sec:method}, we briefly summarize the {\duq} method and describe how we generated the datasets used to train the neural network. Section \ref{sec:results} gives our results for two realistic scenarios. In the first scenario, the network is trained on a complete nuclear mass table from a theoretical calculation; in the second scenario, the network is trained on restricted data given by the Atomic Mass Evaluation \cite{wang2021ame} with the aim of quantifying uncertainties on predictions far away from it.

\section{Theory}
\label{sec:method}

In this section, we briefly introduce the {\duq} method, the construction of training, validation and testing sets, and the neural network architecture.

\subsection{Summary of the {\duq} method}

\begin{figure}[t]
\includegraphics[width=\linewidth]{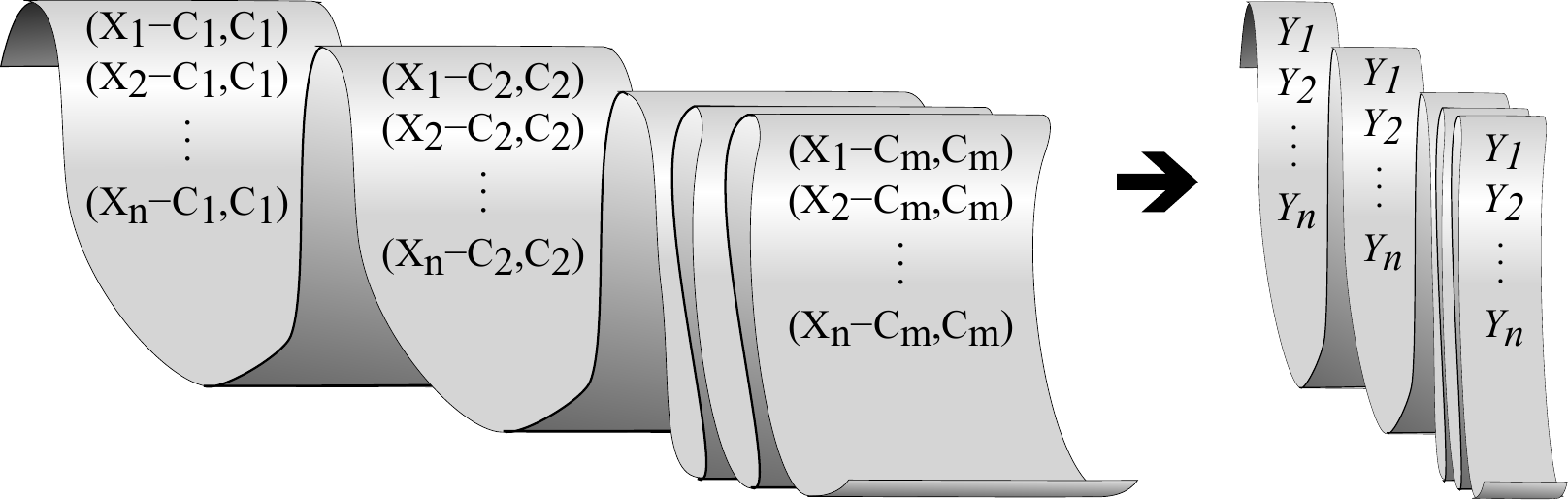}
\caption{{\duq} mapping: each initial input vector $\mlinp_i$ is shifted by the anchor $\anchors_j$. The new input vector is formed by aggregating the shifted vector with the anchor, resulting in input dimensionality $2\diminp$ and a dataset of size $\ntotal \times m$.}
\label{fig:schematic_deltaUQ}
\end{figure}

We use the {\duq} method to estimate the mean and uncertainty of the machine learning predictions \cite{thiagarajan2022single}. The process is as follows. The input feature vectors $\{\mlinp_i\}$ (in our case: $\mlinp_i \propto (N_i, Z_i)$, see Sec.\ref{subsec:data}) are combined with a set of constant bias $\{\anchors_j\}$ called ``anchors'', leading to multiple copies of input combinations $\{\mlinp_i - \anchors_j, \anchors_j\}$, where $i=1,\dots,\dimsamp$ runs through all the input data points and $j$ runs through all the anchors. For convenience, we call the expanded inputs {\duq} anchored inputs. The original input dimension $\diminp$ is thus doubled under the {\duq} scheme and the data length increases from $\dimsamp$ to $\dimsamp\times m$, where $\dimsamp$ is the number of input data points and $m$ is the number of anchors. Figure~\ref{fig:schematic_deltaUQ} illustrates schematically how {\duq} anchored inputs are generated. 

After each epoch, the outputs contain results corresponding to different anchors. The loss is calculated using the mean squared error (MSE) of the results from all the anchors,
\begin{equation}
\loss = \frac{1}{nm}\sum_{i=1}^{n}\sum_{j=1}^{m}(\hat{y}_{ij}-y_i)^2,
\label{eq:loss}
\end{equation}
where $\hat{y}_{ij}$ is the model output for data $i$ with anchor $j$, and $y_i$ is the expected value of output $i$. 

We employ the Adam optimization method \cite{kingma2017adammethodstochasticoptimization}, which adjusts the learning rate automatically in response to gradients calculated during the training. During training, the network parameters are updated to reduce the training loss. The training (validation) loss is calculated by applying \eqref{eq:loss} to the $\ntrain$ ($\nvalid$) points of the training (validation) set and recorded after the network parameters are settled down after each epoch. The optimal network parameters are found when both losses reach the minimum and are stable within certain precision, where we allow for insignificant fluctuations around the true minimum due to the stochastic behavior of the Adam optimization algorithm. 

After the optimal set of parameters of the neural network has been obtained, we compute the predictions. For each data point $i$, the {\duq} method generates a spread of $m$ different values corresponding to the $m$ anchors. One can calculate the mean value of these predictions $\mu_{\Delta}$ for nucleus $i$,
\begin{equation}
\label{eq:mean_duq}
\mu_{\Delta i}=\frac{1}{m}\sum^{m}_j \hat{y}_{ij},
\end{equation}
and the standard deviation $\sigma_{\Delta}$ for nucleus $i$
\begin{equation}
\label{eq:std_duq}
\sigma_{\Delta i}=\sqrt{\frac{1}{m}\sum^{m}_j(\hat{y}^2_{ij}-\mu_{\Delta i}^2)},
\end{equation}
where $i$ runs from 1 to $\ntotal$ for $\ntotal$ being the total number of data points need to be evaluated. 

\subsection{Justification of $\Delta$-UQ}

In this section we give a simple demonstration of why the {\duq} method can estimate epistemic uncertainties.
To simplify our analysis, let us consider only the first linear layer of the neural network because this layer is in direct contact with the {\duq} input transformation. For any given node in this first layer, the original result $y$ is expressed as a function of the input feature $x$ according to $y=kx+b$. With {\duq} the result becomes
\begin{equation}\label{kx+b}
y = (k,\mathscr k)
\begin{pmatrix}
C \\
x - C
\end{pmatrix}+b= \left[\mathscr k+(k-\mathscr k)\frac{C}{x}\right]x+b,
\end{equation}
where $C$ is the {\duq} constant bias, $k$ and $\mathscr k$ are the weights and $b$ is the bias of the linear transformation. Each of these quantities can be multi-dimensional according to the problem. For the sake of simplicity, we neglect $b$ in the current discussion. Training a single node in the first layer of the neural network means finding the optimal number $k$ so that
\begin{equation}
\{kx_1, kx_2, kx_3, ... \}\rightarrow \{y_1,y_2,y_3,...\},
\end{equation}
where $\mlinp=\{x_1,x_2,x_3,...\}$ are the training inputs and $\mlout=\{y_1,y_2,y_3,...\}$ are the targets. In the {\duq} method, the constant biases $C$ are selected from the training data, that is, $\anchors=\{x_1,x_2,x_3,...\}$. We can form the matrix $K$ as
\begin{widetext}
\begin{equation}\label{eq:kk}
K=
\begin{bmatrix}
\mathscr k+(k-\mathscr k)\frac{x_1}{x_1}&\mathscr k+(k-\mathscr k)\frac{x_2}{x_1}&...&\mathscr k+(k-\mathscr k)\frac{x_n}{x_1}\\
\mathscr k+(k-\mathscr k)\frac{x_1}{x_2}&\mathscr k+(k-\mathscr k)\frac{x_2}{x_2}&...&\mathscr k+(k-\mathscr k)\frac{x_n}{x_2}\\
\vdots&\vdots&&\vdots\\
\mathscr k+(k-\mathscr k)\frac{x_1}{x_n}&\mathscr k+(k-\mathscr k)\frac{x_2}{x_n}&...&\mathscr k+(k-\mathscr k)\frac{x_n}{x_n}
\end{bmatrix},
\end{equation}
and training involves finding the mapping
\begin{equation}
K \mlinp \rightarrow \mlout.
\end{equation}

This is still a mapping from $\mlinp$ to $\mlout$. The only change is that the initial weight $k$ (= a single number) becomes a matrix element $K_{ij}$, where $i,j = 1,2,...,n$ and $n$ is the number of training data points. Each $K_{ij}$ explores a different point in the multivariate surface in the hypothesis space at the vicinity of the original solution. It can be seen immediately that one solution is when $n^2$ weights reach an optimum with $k=\mathscr k$. Then $K$ reduces to $n$ identical mappings characterized by $k$ and the neural network has no epistemic uncertainty.

The diagonal terms in \eqref{eq:kk} can be simplified,
\begin{equation}
K =
\begin{bmatrix}
k&\mathscr k+(k-\mathscr k)\frac{x_2}{x_1}&...&\mathscr k+(k-\mathscr k)\frac{x_n}{x_1}\\
\mathscr k+(k-\mathscr k)\frac{x_1}{x_2}& k&...&\mathscr k+(k-\mathscr k)\frac{x_n}{x_2}\\
\vdots&\vdots&&\vdots\\
\mathscr k+(k-\mathscr k)\frac{x_1}{x_n}& \mathscr k+(k-\mathscr k)\frac{x_2}{x_n}&...& k
\end{bmatrix}.
\end{equation}
\end{widetext}
Therefore, if we just select the diagonal terms $K_{ii}=k$, we can go back to $k\mlinp\rightarrow\mlout$. Therefore, $k\mlinp\rightarrow\mlout$ is a subset of solutions of $K\mlinp\rightarrow\mlout$. Thus we prove that {\duq} contains the original solution and it can explore the hypothesis space due to varying weights.

In practice, we can also recover the original solution of $k\mlinp\rightarrow\mlout$ by setting $C=x$ for every data point. Then Eq.~(\ref{kx+b}) becomes
\begin{equation}
y = (k,\mathscr k)
\begin{pmatrix}
x \\
0
\end{pmatrix}+b=kx+b.
\end{equation}
Since $\mathscr k$ is irrelevant, the dimension of the first layer can be reduced by half so that the original neural network structure is recovered.

\subsection{{\duq} and the ensemble method}
\label{app:DUQ}

{\duq} generates epistemic uncertainties that are comparable with the traditional ensemble method; the rigorous proof can be found in the Appendix of Ref.~\cite{thiagarajan2022single}. Here we show a simplified analysis to better understand the connections between the two methods. 

Let us denote \text{std}($\cdot$) as the combination of multiplication of the subsequent layers and computation of the uncertainty. In {\duq}, the estimated uncertainty at the data point $x_1$ is
\begin{multline}
\Delta y_1 = \text{std}\big( kx_1, kx_1 + (k-\mathscr k)(x_2-x_1), \dots, \\
kx_1+(k-\mathscr k)(x_i-x_1), \dots \big)
\end{multline}
that is, 
\begin{multline}
\Delta y_1 = \text{std}\big( kx_1, kx_1 + \Delta k(x_2-x_1), \dots, \\
kx_1 + \Delta k(x_i-x_1),\dots \big),
\label{eq:duq_std}
\end{multline}
respectively. In other words, {\duq} is equivalent to creating an ensemble of $n$ realizations of the same neural network, where the scaling factor $k$ of the first layer of each realization $i$ is obtained by adding a constant random deviation $\Delta k$ scaled by the relative distance of the evaluated point from point $i$. In practice, all inputs and $\anchors$ form a full set of expanded inputs and they are fed to the neural network at each epoch in a single training process.

In the standard ensemble method, the uncertainty at point $x_1$ is simply $\Delta y_1 = \text{std}(kx_1, k_1 x_1, \dots, k_i x_1,\dots)$, where $k_i$, etc., represent different initial weights. This can be recast into
\begin{equation}
\label{eq:ensemble_std}
\Delta y_1  = \text{std}(kx_1, kx_1+\Delta k_1 x_1, \dots, kx_1+\Delta k_i x_1,\dots),
\end{equation}
respectively, with $\Delta k_i = (k_i-k)$. 
Like {\duq}, the ensemble method creates a set of $n$ realizations of the same neural network. However, in contrast to {\duq}, the scaling factor $k$ of the first layer in each realization $i$ is obtained by adding a different random deviation $\Delta k_i$. In the ensemble method, the other layers are also initialized randomly, but in {\duq} they are not affected by the input transformation.

Although {\duq} seems to have some limitation in exploring the hypothesis space, the variations produced after stochastic descent can be similar to the variations generated by the standard ensemble method, especially when the machine learning model approaches a stable minimum in the loss function surface where the vicinity variations in all directions are similar. It turns out that if the machine learning with {\duq} run converges nicely for the training and validation sets, it is nearly impossible for the results from different $C$ to deviate wildly, since the initial weights only differ by a scalar factor. On the contrary, the standard ensemble method trains each copy of the neural network individually, if different runs stop at the same number of epochs, the convergence behavior from each run might differ significantly. Therefore, the {\duq} method generally produces a smaller standard deviation compared to the ensemble method. 

From a practical point of view, {\duq} doubles the input dimension. This change might affect the learning path as well as the convergence rate. Thus, it might not be fair to directly compare the uncertainty predicted by {\duq} and by the ensemble method. Based on that, we focus mainly on obtaining the self-consistency of mean and uncertainty for {\duq}, but not on getting the exact same mean and uncertainty as the standard ensemble method.

\subsection{Datasets Construction}
\label{subsec:data}

Our neural network is trained either on the AME2020 dataset \cite{wang2021ame} (AME) or on a synthetic dataset generated by density functional theory (DFT) calculations. Nuclei computed in DFT include all unstable nuclei predicted to exist between the proton and the neutron dripline. In this work, we will first use the DFT dataset to assess the validity of our method to estimate epistemic uncertainties with {\duq}. We will then simulate a realistic scenario of training on experimental data (as captured by the AME) and making predictions in unknown nuclei.

\subsubsection{DFT calculations}

In DFT, we considered the SLy4 parametrization of the energy functional \cite{chabanat1997skyrme}. 
Since the parametrization of this functional does not specify the pairing channel, we adopted a standard surface-volume, density-dependent pairing force and fitted the pairing strength of both neutrons and protons on the 3-point odd-even mass staggering formula; see \cite{li2024multipole} for additional details. We computed the binding energy of all nuclei with $2\le Z \le 120$ between the proton and neutron dripline, where the neutron (proton) dripline is defined as the set of nuclei where the 2-neutron (proton) separation energy changes sign. The procedure to determine the ground state of all even-even nuclei is described in the Supplemental Material of \cite{navarroperez2022controlling}.
The energy of odd or odd-odd nuclei is computed from blocking calculations in the equal filling approximation \cite{perez-martin2008microscopic}. For both neutrons and protons, the five lowest quasiparticle excitations were considered, thus resulting in 25 different configurations for odd-odd nuclei. The ground-state energy for the odd or odd-odd nucleus is taken as the lowest among all blocking calculations. All calculations were performed with the HFBTHO solver \cite{marevic2022axiallydeformed}. Hereafter, this mass table is referred to as the DFT dataset.

\begin{figure}[t]
\centering
\includegraphics[width=\linewidth]{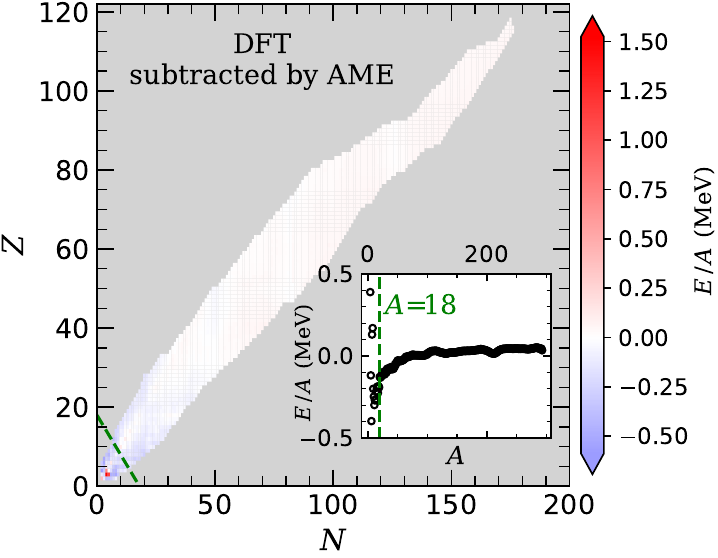}
\caption{The difference between DFT and AME $E/A$ data. The inset shows the average difference calculated by averaging across each isobar.}
\label{fig:differenceDFT2minusAME}
\end{figure}

It is well known that DFT is suboptimal for extremely light nuclei where correlations can be quite large. This is illustrated in Fig.~\ref{fig:differenceDFT2minusAME}, where we show the difference between DFT data and AME data. The inset shows the average difference between DFT and AME data as a function of mass number $A$. Both the landscape plot and the inset show that the difference jumps wildly when $A$ is less than 18. 
Since the quality of the data can affect the training results, we choose to exclude data for nuclei with $A<18$ in the DFT dataset. With this choice, the absolute value of the difference between the DFT dataset and AME data is less than $0.4$ MeV, and the absolute value of the average difference for given $A$ between the DFT dataset and AME data is less than $0.2$ MeV. For a better comparison, we also construct AME dataset with $A\geq 18$.

\subsubsection{Training, validation and testing sets}

\begin{figure}[t]
\centering
\includegraphics[width=\linewidth]{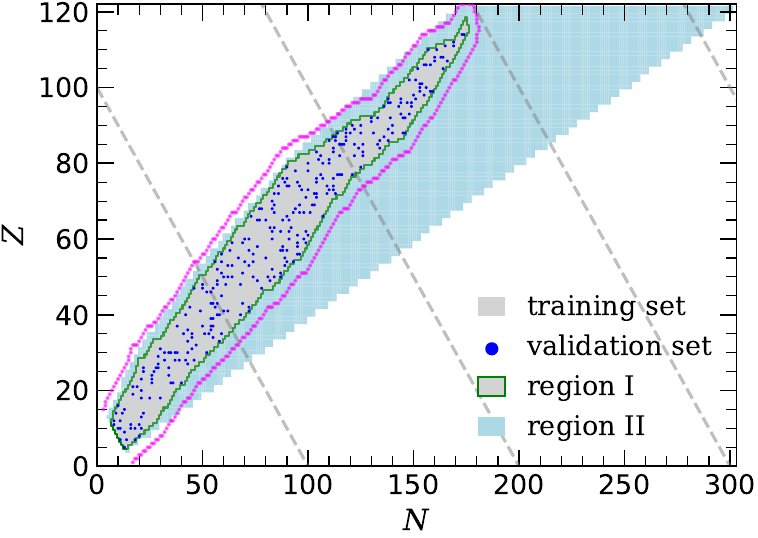}
\caption{Illustration of regions used for training, validation and testing. AME dataset is split into training set and validation set (blue dots) the same way as DFT dataset in region I. DFT dataset in region II forms the testing set of DFT data. The dashed gray lines indicate $A=100, 200, 300, 400$ isobars, whose results will be presented later. The magenta dots indicate the farthest integer points within 5 nuclei away from the boundary nuclei of region I, with $A\geq18$.}
\label{fig:illustrate_DFT_data}
\end{figure}

The size of the AME and DFT datasets is 3099 and 10393, respectively. The AME dataset is randomly split into a training set (90\%) and a validation set (10\%), which contain $\ntrain = 2789$ and $\nvalid = 310$ data points, respectively. Similarly, the DFT dataset in {\em region I}---that coincide with the AME dataset--- is split into the same training and validation sets as the AME dataset, as shown in Fig.~\ref{fig:illustrate_DFT_data}. The DFT dataset in {\em region II} forms the testing set of the DFT data. 

The input features are normalized with Z-score normalization (standardization) to ensure stable and faster optimization, i.e., the input features $\mlinp_i$ are
\begin{equation}
\mlinp_i \equiv (x_{1i},x_{2i})
= \left( 
\frac{N_i-\mu_{N}^t}{\sigma_{N}^t}, \frac{Z_i-\mu_{Z}^t}{\sigma_{Z}^t}
\right), \quad i=1,\dots, n.
\end{equation}
where $\mu_{N}^t$ ($\mu_{Z}^t$) and $\sigma_{N}^t$ ($\sigma_{Z}^t$) are the mean and standard deviation of $N$($Z$) in the training set: $n$ is the number of data points that need to be evaluated. $n=n_t$ in the training phase, $n=n_v$ for validation after each epoch, and $n=n_\text{tot}$ for the evaluation of all data points after the training is complete.

For the training data, we subtract the mean of the binding energy per nucleon $E/A$ from the actual value $E/A$ to reduce the steepness of the sudden decrease of the loss in the first few epochs during the training. The actual output data is thus
\begin{equation}
\mlout_i = \left(\frac{E}{A}\right)_{i} - \mu_{E/A}^t, \quad i=1,\cdots,n,
\end{equation}
where the mean value of the AME (DFT) training set $\mu_{E/A}^t=-8.056$ MeV ($-8.045$ MeV). After training, $(E/A)_i$ is recovered by adding $\mu_{E/A}^t$ back.

\subsection{Network architecture}

The neural network contains four composite layers. Each of the first three composite layers is composed of a linear layer, a nonlinear activation layer made of Sigmoid Linear Unit (SiLU) function, and a batchnorm layer \cite{ioffe2015batchnormalizationacceleratingdeep} to stabilize the variance during training. The last layer consists only of a linear layer. 

We recall that without applying {\duq}, the input dimension is $\diminp = 2$ (for neutron and proton number); with {\duq}, the input dimension is $2\diminp = 4$ (for neutron and proton number and their anchors). The output dimension is $\dimout = 1$ since only the binding energy per nucleon $E/A$ is fit. Each hidden layer contains 32 nodes. Data are fed into the network using multiple mini-batches with 64 data points per mini-batch.
The batchnorm layer normalizes each mini-batch based on their mean and standard deviation before going through the next layer. It regulates the convergence speed regardless of the absolute values of the features. 

We choose initial learning rate to be 0.001 (default) and $\epsilon$ to be 8$\times$10$^{-6}$. $\epsilon$ is a parameter of the Adam algorithm that effectively controls how fast the learning rate adapts to the changing gradient. We set $\epsilon$ to a higher value empirically instead of the default value 10$^{-8}$ to prevent the learning rate from decaying too quickly. In this way, larger patterns of data can be learned before finer adjustments to decrease the loss, so that the learning process is more stable and beneficial to extrapolations.

\begin{figure}[t]
\centering
\includegraphics[width=\linewidth]{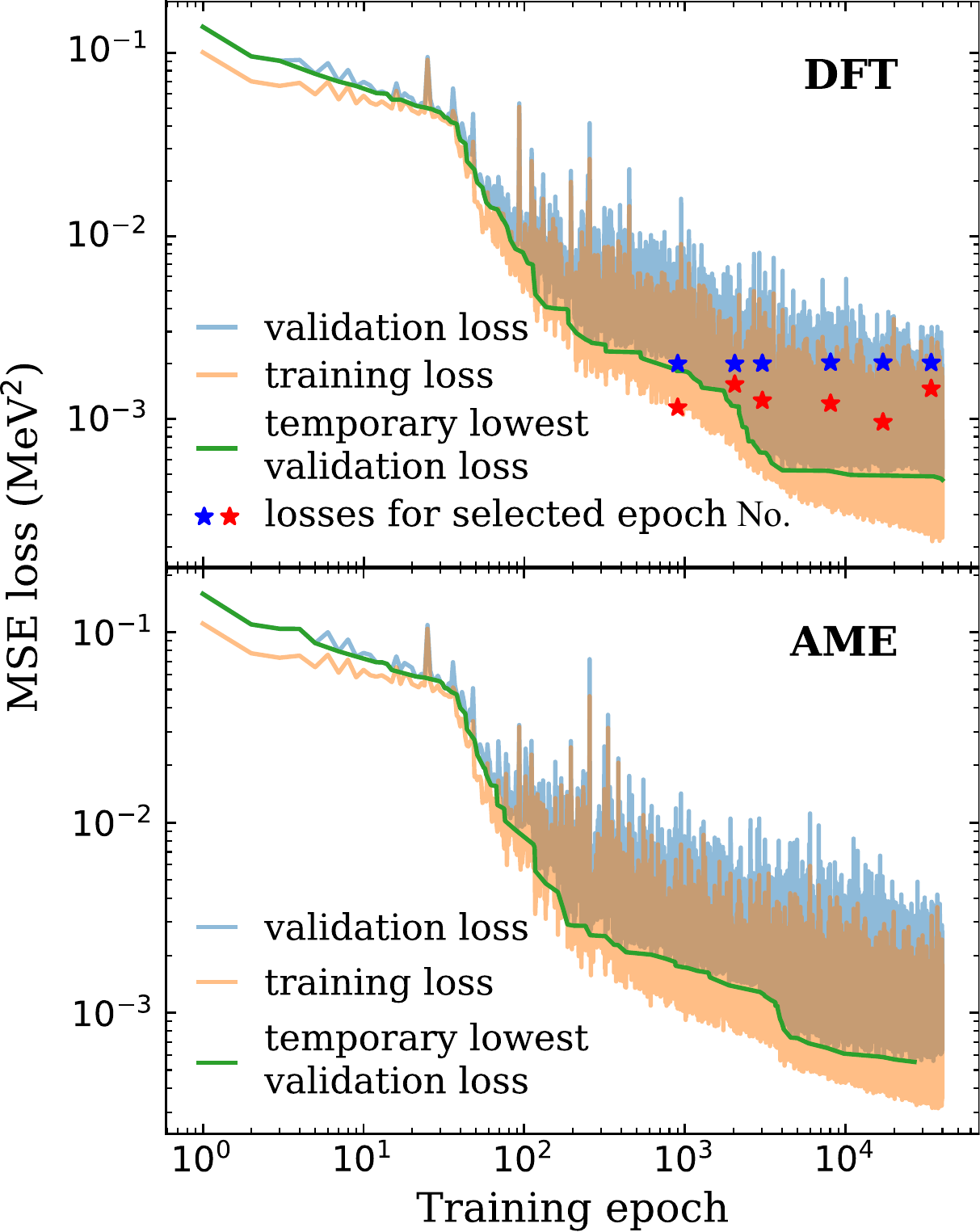}
\caption{Training on DFT or AME data: training and validation MSE loss as a function of the number of epochs. The blue (red) stars are the validation loss (training loss) values for selected epoch number with similar validation loss, which will be discussed later.}
\label{fig:losses_DFT_AME}
\end{figure}

\section{Results}
\label{sec:results}

We first discuss the convergence of the loss function during the training of the networks. Then we illustrate the machine learning results for both DFT and AME with {\duq} uncertainty quantification before comparing the {\duq} uncertainty quantification with the standard ensemble method.

\subsection{The convergence of training process}

Figure \ref{fig:losses_DFT_AME} shows the validation loss and training loss as a function of the number of training epochs when training either on the DFT (top panel) or AME data (bottom panel). The similarities of both loss curves in the first 10$^2$ epochs indicate that the model is learning the same overall trend. After that, the loss curves become noisier and the model is learning the details of the data. We allow the process to run for 40000 epochs and save a copy of the neural network model parameters each time the validation loss becomes lower than at previous epochs. This corresponds to the curve marked ``temporary lowest validation loss'' in both panels. We stop the training process at 40000 epochs because this temporary lowest validation loss up to each epoch flattens out approaching 40000 epochs, as shown in Fig.~\ref{fig:losses_DFT_AME}.

\begin{figure}[b]
\centering
\includegraphics[width=0.95\linewidth]{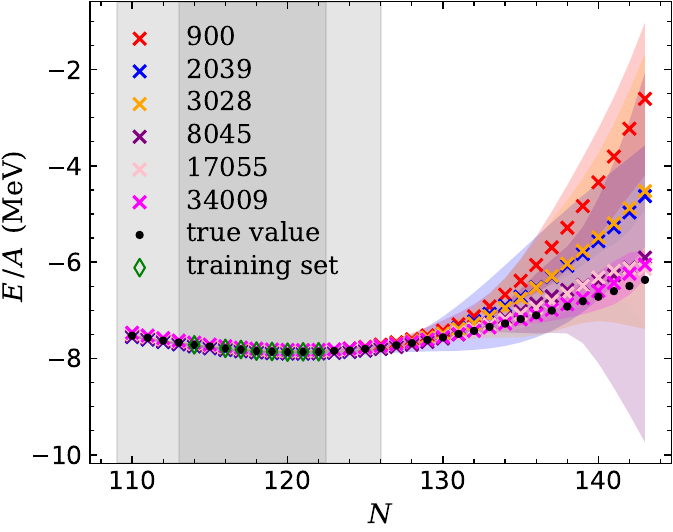}
\caption{Predictions of the binding energy per nucleon $E/A$ for $A=200$ isobars for the selected epochs listed in Table \ref{table_selected_losses}. Crosses represent the mean value and the band indicates three standard deviation estimated with {\duq}.}
\label{fig:convergence_DFT_duq}
\end{figure}

When losses become noisy, the model is still continuously improving. We can see this by examining the results of a few epoch numbers during the training process. Here, we select epoch 900, 2039, 3028, 8045, 17055, 34009. They are chosen so that the difference in their validation losses is less than $10^{-4}$ MeV$^2$, which is negligibly small compared to the loss fluctuations (Table~\ref{table_selected_losses}). However, the predictions are different, as illustrated in Fig.~\ref{fig:convergence_DFT_duq} in the particular case of the isobaric $A=200$ nuclei. 

\begin{table}[!tp]
\caption{Training loss, validation loss and and their difference (MeV$^2$) for training on DFT data at selected epochs.}
\label{table_selected_losses}
\begin{ruledtabular}
\begin{tabular}{rlll}
epoch & validation loss & training loss & difference \\
\hline
 900  & 0.0020015 & 0.0011552 & 0.00084636 \\
 2039  & 0.0020024 & 0.0015453 & 0.00045709 \\
 3028  & 0.0020086 & 0.0012595 & 0.00074908 \\
 8045  & 0.0020343 & 0.0012146 & 0.00081966 \\
 17055  & 0.0020338 & 0.0009605 & 0.00107330 \\
 34009  & 0.0020264 & 0.0014621 & 0.00056429 \\
\end{tabular}
\end{ruledtabular}
\end{table}

It is important to emphasize that a smaller validation or training loss, or a smaller difference between validation and training loss, indicates convergence of the training but not necessarily better results. For example, both the training and validation loss at epoch 900 are smaller than at epoch 2039, but epoch 2039 gives a better extrapolation. Similarly, the difference between validation and training loss at epoch 17055 is the largest in the set, while its predictions are closer to the true value in the extrapolation region. It can be observed that the model captures a new trend each time the temporary lowest validation loss has a significant drop, however, the oscillations above that temporary lowest validation loss will not change the quality of the fit, as long as the temporary lowest validation loss remains the same (until the next significant drop).

In the final plateau of temporary lowest validation loss, models produce very similar fits regardless of the local fluctuations of the loss. The best model is defined as the model that has the lowest validation loss. For DFT, it is obtained at epoch 39858 with training loss 2.46$\times 10^{-4}$ MeV$^2$ and validation loss 4.65$\times 10^{-4}$ MeV$^2$; for AME, the best model is obtained at epoch 26792 with training loss 3.55$\times 10^{-4}$ MeV$^2$ and validation loss 5.51$\times 10^{-4}$ MeV$^2$. 

\begin{figure*}[t]
\centering
\includegraphics[width=0.85\linewidth]{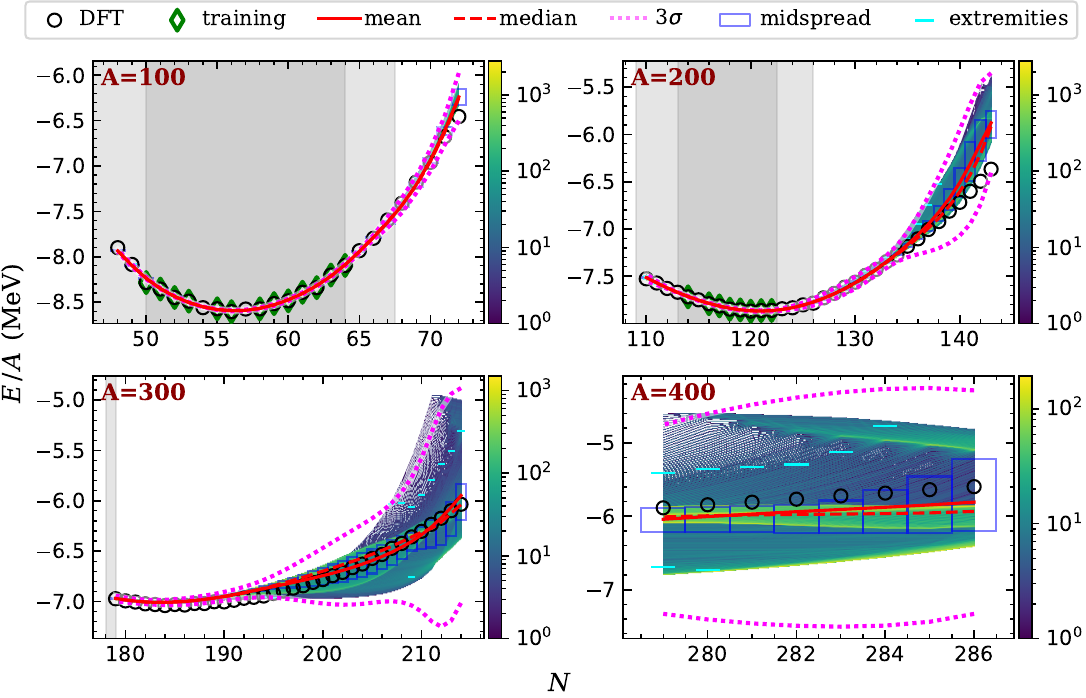}
\caption{$E/A$ from training on DFT with {\duq} uncertainty quantification for selective isobars. See text for explanations.}
\label{fig:DFT_deltaUQ_4}
\end{figure*}

\begin{figure*}[b]
\centering
\includegraphics[width=0.85\linewidth]{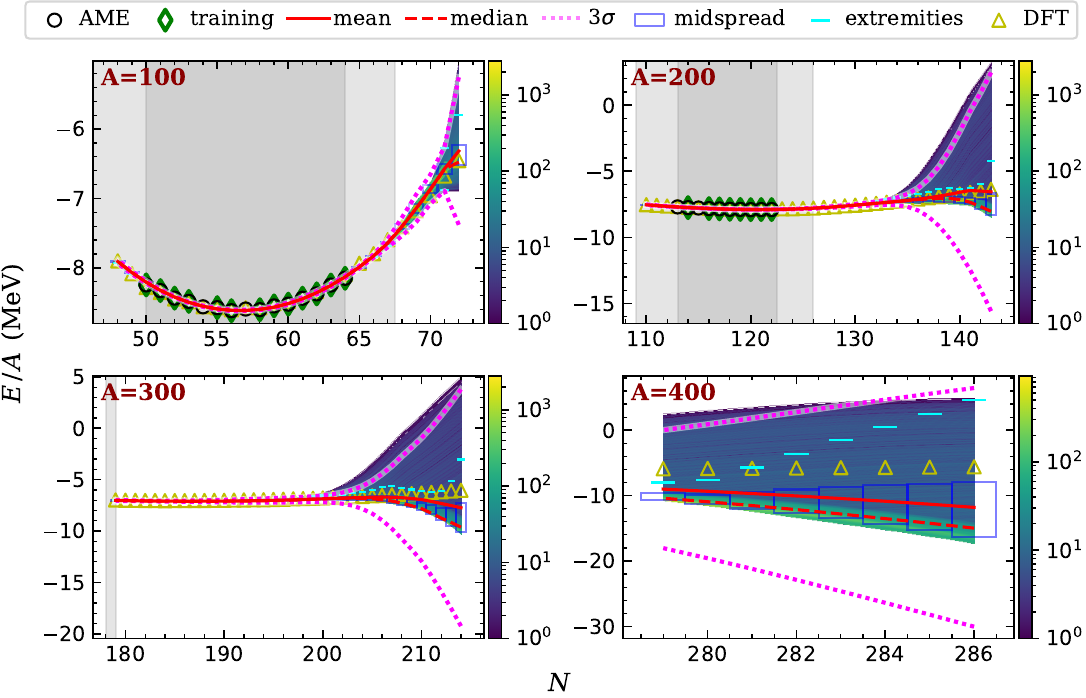}
\caption{Same as Fig.~\ref{fig:DFT_deltaUQ_4}, only from the AME training dataset.}
\label{fig:AME_deltaUQ_4}
\end{figure*}

\subsection{Training results with {\duq} uncertainties}

Figures \ref{fig:DFT_deltaUQ_4} and \ref{fig:AME_deltaUQ_4} show the prediction of the neural network for the binding energy per nucleon for the $A=100$, $A=200$, $A=300$, and $A=400$ isobars with {\duq} uncertainty quantification. These isobars correspond to the four diagonal dashed lines plotted in Fig.~\ref{fig:illustrate_DFT_data} and are representative of the results that we obtained. The training data represents a little more than half the entire set of the $A=100$ isobars; conversely, it represents a little less than the entire set of the $A=200$ isobars. None of the nuclei in the $A=300$ isobars are in the training set, although some of them are not too far from it. Finally, the $A=400$ isobars include nuclei that are very far from the training data. Figure \ref{fig:DFT_deltaUQ_4} is obtained by training the network on the DFT dataset, while Fig.\ref{fig:AME_deltaUQ_4} is obtained by training on the smaller AME dataset.

Each panel in Figs.~\ref{fig:DFT_deltaUQ_4}-\ref{fig:AME_deltaUQ_4} shows the initial data (either DFT or AME) as open circles and the training data as losanges. The mean of the {\duq} prediction is shown as a plain line and computed from Eq.\eqref{eq:mean_duq} where all $n_t = 2789$ anchor nuclei are included. For comparison, we also include the median prediction. We recall that the standard deviation of the {\duq} predictions is given by Eq.\eqref{eq:duq_std} or specifically, Eq.\eqref{eq:std_duq}. For each panel, we choose the $3\sigma$ band to represent the {\duq} uncertainty. In Fig.~\ref{fig:DFT_deltaUQ_4}, the true values of DFT in region II are covered by {\duq} uncertainty.

To better visualize the distribution of the results from all the anchors, the $E/A$ range of each panel is discretized into 200 energy bins. The function $E/A(N)$ of each bin is represented as a colored line, the color of which is given by the number of counts in that bin (in log scale). The deviation of the distribution for each nucleus from a normal distribution is first shown by the difference between the mean and median. As we can see, the difference only shows up when the width of the uncertainty band significantly increases. The fact that all the mean values are within or on the margin of the middle 50$\%$ range (midspread, or interquartile range---IQR) further indicates that the distributions are not very far from normal distributions. Another indicator of deviations from a symmetrical normal distribution is the {\em extremities} calculated by Tukey's 1.5 IQR rule \cite{hoaglin1986performance}. In this rule, the maximum (minimum) is redefined as $Q_3+1.5\times\text{IQR}$ ($Q_1-1.5\times\text{IQR}$) when $Q_3+1.5\times\text{IQR}$ ($Q_1-1.5\times\text{IQR}$) is smaller (larger) than the true maximum (minimum) of the data, where $Q_3$ ($Q_1$) is the third (first) quartile of the data. And any data outside of the redefined maximum and minimum is a statistical outlier. These redefined extremities are shown in the figures. There seem to be more outliers above the maximum than below the minimum. However, these outliers have little effect in skewing the overall distribution because of their relatively lower density.
The conclusion of this discussion is that the $3\sigma$ band originally derived under the assumption of normal distribution remains a reasonable quantity for uncertainty estimation in our case.

Figure \ref{fig:AME_deltaUQ_4} shows that training on AME data gives stable results close to the training data but that the uncertainty explodes very quickly at larger distance. Note that the range of the $y$-axis for the $A=200$, $A=300$ and $A=400$ isobars is considerably larger in Fig.~\ref{fig:AME_deltaUQ_4} than in Fig.~\ref{fig:DFT_deltaUQ_4}. This can be partially attributed to the fact that the noises of DFT data have a consistent bias, as shown in Fig.~\ref{fig:differenceDFT2minusAME}, while the experimental data are more random. This randomness helps machine learning to overcome some local minima and may be the reason why predictions are better close to the training data. However, if we continue to extrapolate further, beyond a certain point, this randomness results in very different extrapolations for different anchors, so the uncertainty band explodes.

\begin{figure*}[t]
\centering
\includegraphics[width=0.85\linewidth]{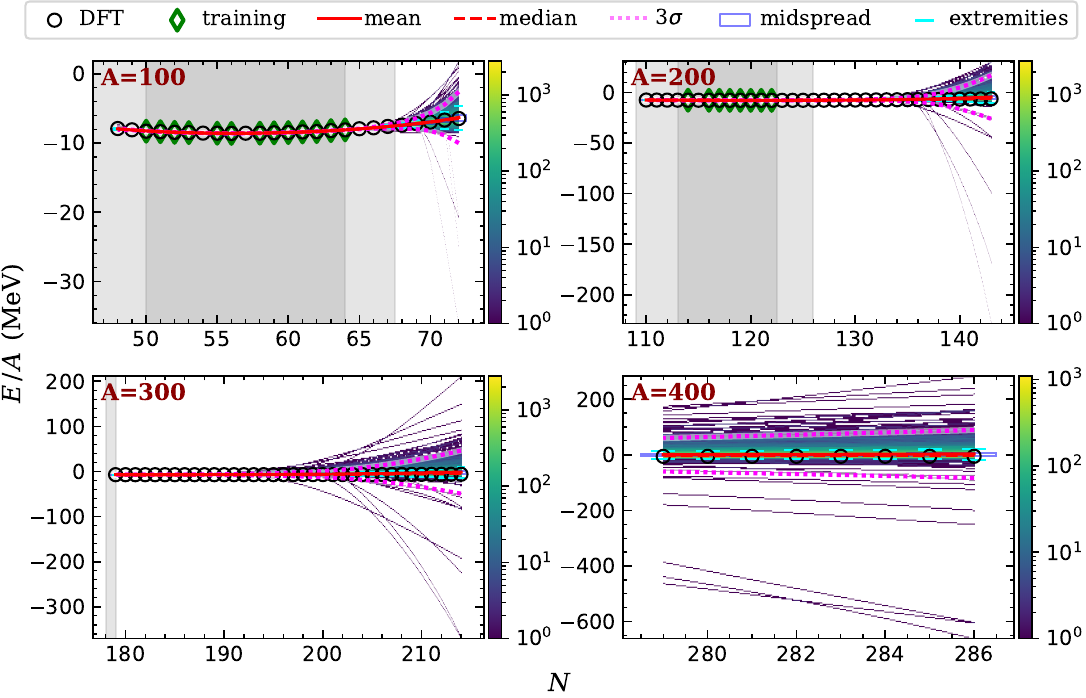}
\caption{$E/A$ from training on DFT with ensemble uncertainty quantification for selective isobars. The color scale is used to visualize the distribution of the results from all the ensemble runs. See text for explanations.}
\label{fig:DFT_ensemble_4}
\end{figure*}

\subsection{Comparison with the ensemble method}

To compare the prediction of uncertainties of {\duq} with the standard technique of ensemble runs, we
performed a set of 2789 runs using the DFT dataset---the exact same number of runs as anchors in the {\duq} method. In each run, the linear layers of the neural network are initialized with a different set of weights using Kaiming uniform \cite{7410480}, which is a scaled version of the uniform distribution popularly used to balance the variance between input and output, by applying a scale factor (Kaiming gain) to account for the variance change related to the input dimension of each linear layer in the deep neural network. The initial biases are also chosen from a uniform distribution scaled to match the magnitude of the output resulting from the Kaiming gain to ensure a stable and effective training process.
The network layer-wise structure and dimensions are the same as the one we used in {\duq} method, except the input of the first linear layer is reduced back to $n_d$ dimensional without requiring the input of anchors. After 40000 epochs, the best model is saved at the lowest validation loss. The results for the selected isobars are shown in Fig.~\ref{fig:DFT_ensemble_4} with the same conventions as Figs.~\ref{fig:DFT_deltaUQ_4}-\ref{fig:AME_deltaUQ_4}. Without {\duq}, individual run takes approximately 2.7 GPU hours on an NVIDIA V100 GPU, and the ensemble of 2789 runs cumulatively consumes 7530 GPU hours. On the other hand, the single {\duq} run that encompasses 2789 anchors takes about 5 (2.4) GPU hours on an NVIDIA V100 (H100) GPU.

\begin{figure}[b]
\centering
\includegraphics[width=0.95\linewidth]{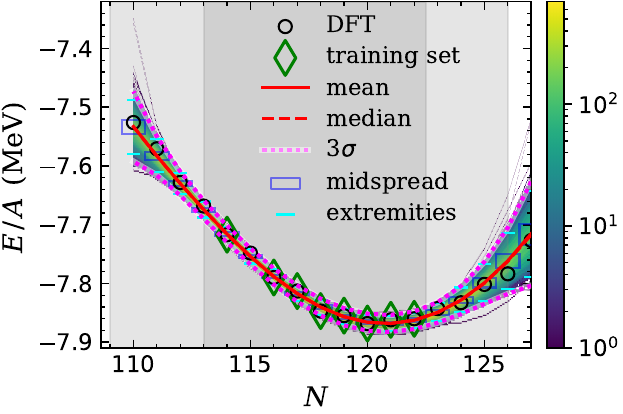}
\caption{A zoom-in figure of the ensemble method for $A=200$ and $N<127$, with the color-mapped bins adjusted for the $E/A$ range in this window.}
\label{fig:DFT_ensemble_A200}
\end{figure}

As expected, some of the ensemble runs produce exceptionally large deviations in the extrapolation region that are visible as additional colored lines in each panel. Even though the tipping point of the explosion in uncertainty, around $N=134$ in $A=200$ case for example, is the same in both {\duq} (Fig.~\ref{fig:DFT_deltaUQ_4}) and ensemble method (Fig.~\ref{fig:DFT_ensemble_4}), the increase in uncertainty in the ensemble method is considerably larger. The uncertainty band of the ensemble method can be one or two magnitudes larger than {\duq}. Note that each run, including the ones giving these very large deviations, still gives a good reproduction of the {\em training} data, as can be seen in Fig.~\ref{fig:DFT_ensemble_A200}, which shows the predictions for the $A=200$ isobaric line only between $N=110$ and $N=126$.
In other words, the ensemble method greatly overestimates the epistemic uncertainty of the network since it does not provide any quality control and does not account for the quality of the local minimum the network discovers during training. 
On the other hand, {\duq} method uncertainty represents the epistemic uncertainty of a good local minimum.

\begin{figure}
\centering
\includegraphics[width=\linewidth]{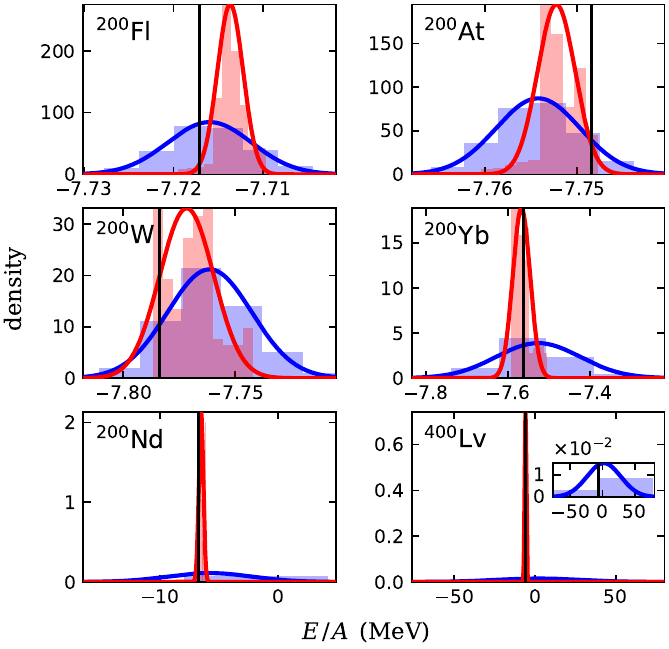}
\caption{Density distribution of the results from {\duq} (red) and the ensemble method (blue) for selected nuclei, for training on the DFT data. A Gaussian fit curve is shown along with each histogram. To clearly illustrate the fit, the figure is adjusted to the window of three standard deviations of the ensemble method results, with the vertical range adjusted to the maximum of both {\duq} and the ensemble Gaussian fit. The inset of the last figure is a zoom-in figure for the Gaussian fit of the ensemble results. The vertical black line is the true value. }
\label{fig:DFT_selected_nuclei}
\end{figure}

To get a better representation of this effect, we show in  Fig.~\ref{fig:DFT_selected_nuclei} the density distribution of the predictions for selected nuclei. For each nucleus, we plot the density distribution extracted from the {\duq} calculations and from the ensemble runs, with selected nuclei as examples. Five nuclei are chosen from $A=200$ isobars, with $N=114$ ($^{200}$Fl), $N=115$ ($^{200}$At), $N=126$ ($^{200}$W), $N=130$ ($^{200}$Yb), and $N=140$ ($^{200}$Nd). One additional nucleus from the $A=400$ isobar, $^{400}$Lv ($N$=284), is also included. 

$^{200}$Fl and $^{200}$At are included in the training and validation set, respectively, while $^{200}$W is five nuclei away from the nearest training data in the testing set. In these three nuclei, {\duq} and the ensemble method predict comparable uncertainties, with the ensemble method uncertainty being slightly larger. 
By contrast, $^{200}$Yb is eight nuclei away from the nearest training data along the same isobar, but the uncertainty is still relatively small, with the ensemble method uncertainty being around three times of the {\duq} uncertainty. Both $^{200}$Nd and $^{400}$Lv are far from any training data in the region where the epistemic uncertainty increases significantly. In these two nuclei, the ensemble method uncertainty is in fact orders of magnitude larger than what {\duq} predicts.

\section{Conclusion}\label{conclusion}
\label{sec:conclusion}

In this paper, we apply the {\duq} method to assess the epistemic uncertainty of machine learning models of the nuclear binding energy, which can be thought of as a simpler prototype of more complex deep-learning problems. We make use of two sets of data, one is calculated by density functional theory (DFT), the other is the AME2020 compilation. The AME2020 dataset is used to define the region (called ``region I'') for training and validation for both AME2020 and DFT datasets. The DFT dataset has an additional testing set outside of the AME2020 region (called ``region II''), which enables us to examine the reliability of the machine learning with {\duq} method when extrapolating far from stability. 

We show that {\duq} gives accurate estimates of the uncertainty band, which follows the expected trend as its width widens when the evaluated nucleus is far from the training set nuclei. Furthermore, {\duq} ability to signal when machine learning results become unreliable as we extrapolate outward from the training region is superior to ensemble methods. %
The lack of fine-tuning control for the results produced by the thousands of individual ensemble runs can cause non-realistic, very large deviations in the extrapolation region. Therefore, ensemble methods can only estimate the upper bound of epistemic uncertainty. Instead, {\duq} estimates the epistemic uncertainty around the best fit, after all the machine learning parameters have been carefully tuned.
In addition, {\duq} is flexible in that it can be easily implemented to any deterministic or probabilistic neural network, by expanding the input dimension of the first linear layer. It can be used either for epistemic uncertainty quantification or simply to test the stability of training results responsive to changing initial weights. In contrast to the multiple runs needed by ensemble method to obtain epistemic uncertainties, {\duq} only needs one run, thereby significantly reducing the amount of computational resources.

\begin{acknowledgments}
This work was performed under the auspices of the U.S. Department of Energy by Lawrence Livermore National Laboratory under Contract DE-AC52-07NA27344. This material is partially based upon work supported by the U.S. Department of Energy, Office of Science, Office of Advanced Scientific Computing Research and Office of Nuclear Physics, Scientific Discovery through Advanced Computing (SciDAC) program. Computing support for this work came from the Lawrence Livermore National Laboratory Institutional Computing Grand Challenge program. We are especially grateful to Jayaraman J. Thiagarajan and Vivek Narayanaswamy, co-inventors of the {\duq} method, for fruitful discussions.
\end{acknowledgments}

\appendix

\vspace{0.7cm}

%


\end{document}